\documentclass{aa}
\usepackage{graphicx,amsmath,amssymb}
\begin{document}
\def\be{\begin{equation}}
\def\ee{\end{equation}}
\def\ie{{\it i.e.}}
\def\eps{\varepsilon}
\def\etl{{\it et al. }}
\def\F{{\rm F} }
\def\d{{\rm d}}
   \title{Non-equilibrium beta processes in superfluid neutron star cores}

   \author{L. Villain \inst{1,2} \and P. Haensel \inst{1}}
\institute{N. Copernicus Astronomical Center, Polish Academy of Sciences,
Bartycka 18, PL-00-716 Warszawa, Poland
\and LUTH, UMR 8102 du CNRS, Observatoire de Paris-Meudon, F-92195 Meudon Cedex, France}
\offprints{L. Villain \email{loic@camk.edu.pl}}

   \abstract{The influence of nucleons superfluidity on the beta
   relaxation time of degenerate neutron star cores, composed of
   neutrons, protons and electrons, is investigated. We numerically
   calculate the implied reduction factors for both direct and
   modified Urca reactions, with isotropic pairing of protons or
   anisotropic pairing of neutrons. We find that due to the non-zero
   value of the temperature and/or to the vanishing of anisotropic
   gaps in some directions of the phase-space, superfluidity does not
   always completely inhibit beta relaxation, allowing for some
   reactions if the superfluid gap amplitude is not too large in
   respect to both the typical thermal energy and the chemical
   potential mismatch. We even observe that if the ratio
   between the critical temperature and the actual temperature is very
   small, a suprathermal regime is reached for which superfluidity is
   almost irrelevant. On the contrary, if the gap is large enough, the
   composition of the nuclear matter can stay frozen for very long
   durations, unless the departure from beta equilibrium is at least
   as important as the gap amplitude. These results are crucial for
   precise estimation of the superfluidity effect on the
   cooling/slowing-down of pulsars and we provide online subroutines
   to be implemented in codes for simulating such evolutions.
    }
   \titlerunning{Non-equilibrium beta processes in superfluid NSs}
   \authorrunning{L. Villain \& P. Haensel}

   \maketitle

%%%%%%%%%%%%%%%%%%%%%%%%%%%%%%%%%%%%%%%%%%%%%%%%%%%%%%%%%%%%%%
\section{Introduction}
\label{sect:introduction}
%%%%%%%%%%%%%%%%%%%%%%%%%%%%%%%%%%%

In the simplest model, the so-called $npe$ matter model, neutron star
(NS) cores are transparent for neutrinos and mainly consist of
degenerate neutrons, with a small admixture of equal numbers of
protons and electrons. In such conditions, the composition is
characterized by the fraction of protons among the total number of
nucleons, $x_p$. Due to beta reactions, the value of $x_p$ depends on
density and is usually determined by the condition of beta
equilibrium, for instance in the calculations of non-dynamical
quantities, such as the equation of state (EOS). However, the
assumption of beta equilibrium is only valid if the density of a
matter element changes on a timescale $\tau_\rho$ much longer than the
beta equilibration timescale $\tau_\beta$. But in strongly degenerate
$npe$ matter, $\tau_\beta$ is made much longer than in normal matter
by the decrease of the available phase-space due to the Pauli
exclusion principle. As a consequence, various astrophysical scenarios
in which the condition $\tau_\beta\ll \tau_\rho$ is violated were
pointed out by several authors. Such a situation occurs in
gravitational collapse of neutron star (Haensel \cite{h92},
Gourgoulhon \& Haensel \cite{gh93}), during pulsar spin-down
(Reisenegger \cite{reisenegger95}, \cite{reisenegger97}, Fern\'andez
\& Reisenegger \cite{fernandez05}), or in the neutron star interiors
just due to the existence of even relatively slow hydrodynamic flows
(for an example, see Urpin \& Shalybkov \cite{UrpSha1996}) or of
millisecond oscillations (Reisenegger \& Goldreich
\cite{reisenegger92}, Villain et al. \cite{villain05}).\\

Nonetheless, beta equilibrium breaking and change of composition are
crucial to take into account not only from those mainly ``mechanical''
points of view, but also in the thermal evolution of not too old neither
too young NSs. Indeed, the cooling of such NSs proceeds through the
emission of neutrinos by the cold (\ie{}, degenerate) $npe$ matter,
whose deviation from beta equilibrium is characterized by the chemical
potential mismatch
\mbox{$\delta\mu\,\equiv\,\mu_n-\mu_p-\mu_p$}. While in beta
equilibrium only the non-zero value of the temperature $T$ determines
the size of the available phase-space, and consequently all reactions
rates, \mbox{$\delta\mu\neq 0$} opens additional volume in the
phase-space for beta processes. This results in an increase of the
neutrino emissivity, and as beta reactions occurring off-equilibrium
produce entropy and therefore heat neutron star matter, the net effect
is the matter heating (Haensel \cite{h92}, Gourgoulhon \& Haensel
\cite{gh93}, Reisenegger \cite{reisenegger95},
\cite{reisenegger97}, Fern\'andez \& Reisenegger \cite{fernandez05}).\\

 Furthermore, the description of any basic physical process that takes
place inside NSs can be made more difficult by another phenomenon, which
is the possible superfluidity of nucleons. Indeed, it is expected that
in about one year after neutron star birth, neutrons (protons) in the
core become superfluid, when temperature goes below critical one,
\mbox{$T_{{\rm c}n}$} (\mbox{$T_{{\rm c}p}$}). Then, energy gaps,
$\Delta_n$ and $\Delta_p$, appear in the nucleon excitation energy
spectra, reducing again the available phase-space. This strongly slows
down the beta processes in dense matter for \mbox{$T\ll T_{\rm c}$}
and also makes the dynamical properties, such as pulsations (Lee
\cite{lee95}, Andersson \& Comer \cite{andersson01}, Prix \& Rieutord
\cite{prix02}), more complicated to evaluate.\\

 To summarize, the beta processes rate, in its full complexity,
involves at least four parameters with the dimension of energy:
\mbox{$\delta\mu$, $k_{\rm B}T$, $\Delta_n$, and $\Delta_p$}. The
roles of the first and the second pairs of parameters are opposite:
\mbox{$\delta\mu$, $k_{\rm B}T$} increase the available phase-space,
while the superfluid energy gaps inhibit the reactions. In the limit
of small deviations from the thermodynamic equilibrium,
\mbox{$\delta\mu/k_{\rm B}T\ll 1$}, the beta relaxation rate can be
linearized in this small parameter and the dissipation can be
represented by the bulk viscosity (Haensel \etl \cite{hly00},
\cite{hly01}, \cite{hly02}, and references therein). Nevertheless, in
the general case, both the heating rate and neutrino losses have to be
calculated explicitly in the out of beta equilibrium $npe$ matter. For
a non-superfluid matter this was done in Haensel (\cite{h92}),
Gourgoulhon \& Haensel (\cite{gh93}), Reisenegger
(\cite{reisenegger95}) and Fern\'andez \& Reisenegger
(\cite{fernandez05}). The effect of nucleon superfluidity on beta
relaxation rates was studied using a very crude model by Reisenegger
(\cite{reisenegger97}). He considered deviation from beta equilibrium
implied by the the pulsar spin-down, and assumed that for
\mbox{$\delta\mu<\Delta_p+\Delta_n$} the beta reactions were
completely blocked (no reactions), while for
\mbox{$\delta\mu>\Delta_p+\Delta_n$} the effect of superfluidity could
be neglected (normal matter rates). As we should discuss in more
details later, such a step-like modeling of the reduction factor
behaviour is unrealistic.\\

In the present article, we calculate numerically the superfluid
reduction factors for a broad range of the relevant parameters making
also available on-line subroutines to proceed to any farther
calculation. Both direct Urca and modified Urca processes are
considered in the presence of various types of superfluidity. In this way, we get
correct reduction factors that can be used for simulating the
evolution of superfluid neutron star cores which are off beta
equilibrium.\\

The formulae for the rates of non-equilibrium beta processes in a
non-superfluid $npe$ matter are reminded in Sect.\
\ref{sect:nonsup.beta}. Basic features of nucleon superfluidity in NSs
are presented in Sect.\ \ref{sect:sup.nuc}. In Sect.\ \ref{sect:R.np},
the formulae for the superfluid reduction factors are derived, with
some illustrative numerical results in Sect.\ \ref{sect:R.results}. In
Sect.\ \ref{sect:disc.concl}, we discuss our results and their
possible application in numerical simulations of neutron star
evolution and dynamics. Finally, in the Appendix, the subroutines to
calculate the reduction factors for such numerical simulations are
briefly described, together with some comments on their practical use
and the address of the website from which they can be downloaded.

%%%%%%%%%%%%%%%%%%%%%%%%%%%%%%%%%%%%%%%%%%%%%%%%%%%%%%%%%%%%%%
\section{Out-of-equilibrium beta processes without superfluidity}
\label{sect:nonsup.beta}
%%%%%%%%%%%%%%%%%%%%%%%%%%%%%%%%%%%

Non-equilibrium beta processes in normal (non superfluid) neutron star
cores have already been the subject of several studies (Haensel
\cite{h92}, Reisenegger \cite{reisenegger95}). In this section, we
shall not discuss the astrophysical conditions in which the breaking
of chemical equilibrium can occur, but focus on its microphysical
description. Moreover, we just give here a brief summary of known
results, introducing our notations in units $\hbar=k_{\rm B}=c=1$
(with exception of Sect. \ref{sect:sup.nuc}), mainly following the
conventions of Haensel (\cite{h92}).\\

We take the basic $npe$ model of nuclear NS matter and assume that
\begin{itemize}
\item [-] the core's content is in a stationary state (no
oscillation), with all particles comoving. The outcome of a possible
difference between the motions (e.g., rotation) of $n$ and $p$ fluids
(a crucial phenomenon for superfluids) will be considered elsewhere;
\item [-] the nucleons are strongly interacting non-relativistic Fermi
liquids and the electrons form an ultrarelativistic ideal Fermi
gas. The temperature is sufficiently low for the $npe$ components
to be strongly degenerate, and we shall then work for all of them with
quantities at zero temperature and at thermodynamic, but not
necessarily chemical, equilibrium: beta equilibrium is not
assumed. Hence, the Fermi momenta $p_{\F i}$, for the $i$ species,
versus the respective densities $n_i$, are \mbox{$p_{\F i}= \left(3 \pi^2
n_i \right)^{1/3}$};
\item [-] local electromagnetic equilibrium is reached and matter is
electrically neutral;
\item [-] the neutrinos freely escape from the star.
\end{itemize}

Cooling of NSs involves various reactions, among which the Urca
processes are the only ones that we should deal with in this
article. Indeed, as neutrinos freely escape from the star, other
reactions, such as bremsstrahlung, do not change the chemical
composition of $npe$ matter and are then not affected by beta
equilibrium breaking. Yet, a distinction has to be made between two
types of Urca processes, since the direct Urca reactions (called
thereafter Durca, following K. Levenfish, see, Yakovlev \etl
\cite{ykgh01}) are kinematically allowed only if the proton fraction
is high enough. If not, the modified Urca reactions
(Murca)\footnote{with, in respect to Durca, an additional spectator
nucleon.} play the key-role, nonetheless they can be neglected when
Durca occur. From a practical point of view, it means that there is a
threshold density below which one has to deal with (and only with)
Murca reactions, while, for densities above that value, it is
sufficient to consider only Durca processes. As the latter are the
simplest, we shall start with their description.

\subsection{Direct Urca processes}

In $npe$ matter, Durca reactions are allowed if
\mbox{$p_{\F n}\,<\,p_{\F p}\,+\,p_{\F e}$}. Assuming local neutrality
of matter, it is equivalent to \mbox{$p_{\F n}\,<\,2\,p_{\F p}$} and
then to $n_p\,>\,n/9$, where $n$ is the total baryonic density
\mbox{$n\,=\,n_p\,+\,n_n$}. This leads to a threshold value of the
density that depends on the equation of state, and mainly on the
symmetry energy, but which is typically several times the standard nuclear
matter density \mbox{$n_0\,=\,2.8\,\times 10^{14} {\rm g}\,{\rm cm}^{-3}$}.\\

When beta equilibrium is assumed, the two Durca reactions,
\be \label{Eq:Durca}
\begin{split} 
D_n:&\hspace{1cm}n\,\to\,p\,+\,e^-\,+\,\bar{\nu}_e\,,\\
D_p:&\hspace{1cm}p\,+\,e^-\,\to\,n\,+\,\nu_e\,,
\end{split}
\ee have the same rates and form, all together, the usual Durca
process. As we do not assume beta equilibrium, we have to consider two
different rates (numbers of reactions per ${\rm cm}^3$ and during one
second) (see Haensel \cite{h92}, Reisenegger \cite{reisenegger95})
\begin{multline} \label{Eq:Gamma_dn}
\Gamma_{D_n} \,=\,  \int
        \frac{\d \vec{p}_e}{(2\pi)^3 }     \,
        \frac{\d \vec{p}_p}{(2\pi)^3 }     \,
        \frac{\d \vec{p}_n}{(2\pi)^3 }     \,
        \frac{\d \vec{p}_\nu}{(2\pi)^3 }\,(1 -f_e)\, (1 -f_p) \,f_n \\
           \times \, (2\pi)^4\, \delta(E_f - E_i) \,
          \delta(\vec{P}_f - \vec{P}_i)\, |M_{if}|^2
\end{multline}
and
\begin{multline} \label{Eq:Gamma_dp}
\Gamma_{D_p} \,=\,  \int
        \frac{\d \vec{p}_e}{(2\pi)^3 }     \,
        \frac{\d \vec{p}_p}{(2\pi)^3 }     \,
        \frac{\d \vec{p}_n}{(2\pi)^3 }     \,
        \frac{\d \vec{p}_\nu}{(2\pi)^3 }\,(1 -f_n)\,f_p \,f_e\\
           \times \, (2\pi)^4\, \delta(E_f - E_i) \,
          \delta(\vec{P}_f - \vec{P}_i)\, |M_{if}|^2\,.
\end{multline}

In these equations, $|M_{if}|^2$ is the squared transition amplitude
summed and averaged on spins states, \mbox{$f_i\,\widehat{=}\,
(1\,+\,e^{(\eps_i-\mu_i)/T})^{-1}$} is the \mbox{Fermi-Dirac}
distribution for a fermion $i$ with energy $\eps_i$, chemical
potential $\mu_i$, and temperature $T$, while the $\delta$-functions
ensure the conservation of the total energy and momentum. The physical
quantity we will focus on is $\Delta \Gamma_D$, that we define as the
difference between $\Gamma_{D_n}$ and $\Gamma_{D_p}$: \mbox{$\Delta
\Gamma_D\,\widehat{=}\,\Gamma_{D_n}-\Gamma_{D_p}$}. This quantity
characterizes the relaxation rate (and hence timescale) for reaching beta equilibrium.\\

In the context of strongly degenerate matter, there is a standard
approximation, that can be applied here, to go beyond the previous
formulae in an analytical way, the so-called phase-space decomposition
(see Shapiro \& Teukolsky \cite{st83}). It mainly consists in
replacing in the integrals all smooth functions by their values at
\mbox{$p_i\,=\,p_{\F i}$}, which enables to factorise out from the
integrals the mean value of the microphysical factor $|M_{if}|^2$. We
shall not make explicit the whole calculation, and send the
reader to Yakovlev \etl (\cite{ykgh01}) for a detailed review. However, with the
following dimensionless variables ($i=n, p, e$)
\begin{equation} \label{Eq:xdef1}
   x_i\,=\,\frac{\eps_i- \mu_i}{T}, \quad
   x_\nu\,=\,\frac{\eps_\nu}{T}, \quad
   \xi\, =\, \frac{\delta \mu}{T},
\end{equation}
where we defined \mbox{$\delta \mu\,\widehat{=}\,\mu_n\,-\,\mu_p\,-\,\mu_e$}, we finally get
\begin{equation} \label{Eq:detgamdint}
\Delta \Gamma_D(\xi)\,=\,\Delta \Gamma_{D_0}\, I_D(\xi)\,,
\end{equation}
with the dimensionless integral
\begin{multline} \label{Eq:ID}
I_D(\xi) \, = \, \int\limits_0\limits^\infty \d x_\nu \; x_\nu^2
       \int\limits_{-\infty}\limits^{+\infty} \d x_n\; \d x_p\; \d x_e \\
 \{f_n\,(1-f_p)\,(1-f_e)\,\delta \left(x_n-x_p-x_e-x_\nu+\xi \right)
 \\
- f_p\,f_e\,(1-f_n)\,\delta \left(x_n+x_\nu-x_p-x_e+\xi \right)\}\,.
\end{multline}
The rate $\Delta \Gamma_{D_0}$, a physical factor that depends on the
EOS, takes such typical values that the relaxation time, once the
integral (\ref{Eq:ID}) calculated, is around
\mbox{$\tau_{\rm rel}^{(D)}\,\sim\,20\,T_9^{-4}$ s}, with
$T_9\,=\,T/(10^9\,{\rm K})$ (Yakovlev \etl \cite{ykgh01}).\\

As far as the calculation of the integral (\ref{Eq:ID}) is concerned,
using ${\rm f}(x_i)\,\widehat{=}\,f_i\,=\,1\,-\,{\rm f}(-x_i)$
combined with known results on Fermi integrals, we are lead to
\begin{equation} \label{Eq:ID2}
I_D(\xi)\, =\,  \int\limits_0\limits^\infty \d x_\nu \; x_\nu^2\; \{{\rm J_D}(x_\nu-\xi)\,-
\,{\rm J_D}(x_\nu+\xi)\}\,,
\end{equation}
where \mbox{${\rm J_D}(x)\,\widehat{=}\,{\rm
f}(x)\,(x^2+\pi^2)/2$}. The latter integration can now be done
analytically and gives (Reisenegger \cite{reisenegger95})
\begin{equation} \label{Eq:ID3}
I_D(\xi)\,=\,\xi\,\pi^4\,\frac{17}{60}\,\left(1\;+\;\frac{10\;\xi^2}{17\;\pi^2}
+\;\frac{\xi^4}{17\;\pi^4}\right)\,.
\end{equation}

\subsection{Modified Urca}

Durca processes are by many orders of magnitude the fastest beta
processes in NS cores. However, as already mentioned, they can be
kinematically forbidden. If this is the case, the Murca reactions
prevail, which involve an additional spectator nucleon to ensure the
conservation of both energy and momentum. This additional nucleon can
be either a proton or a neutron. Hence, the distinction has to be made
between two branches, and not only when some nucleons are
superfluid. The Murca reactions are \be \label{Eq:Murca}
\begin{split}
M^N_n:&\hspace{1cm}n\, +\, N\, \to\, p\, +\, N\, +\, e^-\, +\, \bar{\nu}_e\\
M^N_p:&\hspace{1cm}p\, +\, N\, +\, e^-\, \to\, n\, +\, N\, +\, \nu_e\,,
\end{split}
\ee where $N$ is the spectator nucleon. If this particle is a proton,
these reactions are called {\it proton branch} of Murca, and they are
called {\it neutron branch} whether it is a neutron\footnote{Notice
that the proton branch has a density threshold, but it is much smaller
than the threshold for Durca, with the condition $n_p\,>\,n/65$ in
$npe$ matter (Yakovlev \& Levenfish \cite{yl95}).}.\\

If none of the nucleons is superfluid, we do not really have to make
distinction between the two branches in the phase-space
integral, and using the same notations as in the Durca case (skipping
the details of the calculations that are very similar), we can write
\begin{equation} \label{Eq:detgammint}
\Delta \Gamma_M(\xi)\, =\,\Delta \Gamma_{M_0}\, I_M(\xi)
\end{equation}
with
\begin{multline} \label{Eq:IM}
I_M(\xi) \, = \, \int\limits_0\limits^\infty \d x_\nu \; x_\nu^2
       \int\limits_{-\infty}\limits^{+\infty} \d x_n\; \d x_p\; \d x_e\; \d x_{Ni}\;
\d x_{Nf}\;\\
\{f_n\,f_{Ni}(1-f_p)\,(1-f_e)\,(1-f_{Nf})\\
\times\,\delta \left(x_n+x_{Ni}-x_p-x_e-x_{Nf}-x_\nu+\xi \right)  \\
- f_p\,f_e\,f_{Ni}\,(1-f_n)\,(1-f_{Nf})\\
\times\,\delta\left(x_n+x_{Nf}+x_\nu-x_p-x_{Ni}-x_e+\xi \right)\}\,.
\end{multline}

Here, $x_{Ni}$ and $x_{Nf}$ are respectively the ``$x$ variables'',
defined as in Eq.(\ref{Eq:xdef1}), for the {\bf i}nitial and {\bf
f}inal spectator nucleons. Finally, $\Delta \Gamma_{M_0}$ is a
constant that depends on the EOS and takes such values that the
relaxation timescale is \mbox{$\tau_{\rm rel}^{(M)}\,\sim\,T_9^{-6}$ months}.\\

Equation(\ref{Eq:IM}) can also be written
\begin{equation} \label{Eq:IM2}
I_M(\xi) \,=\,  \int\limits_0\limits^\infty \d x_\nu \; x_\nu^2\; \{{\rm J_M}(x_\nu-\xi)\,-
\,{\rm J_M}(x_\nu+\xi)\}
\end{equation}
where \mbox{${\rm J_M}(x)\,\widehat{=}\,{\rm
f}(x)\,(x^4+10\,\pi^2\,x^2+9\,\pi^4)/24$}, which leads to (Reisenegger
\cite{reisenegger95})
\begin{equation} \label{Eq:IM3}
I_M(\xi)\,=\,\frac{367\;\xi\;\pi^6}{1512}\,\left(1\;+\;\frac{189\;\xi^2}
{367\;\pi^2}
+\;\frac{21\;\xi^4}{367\;\pi^4}+\;\frac{3\;\xi^6}{1835\;\pi^6}\right)\,.
\end{equation}

In the presence of superfluidity, the rates can be much more
complicated to calculate. Indeed, in all the previous calculations, we
could quite easily get rid of all angular integrals as they always
gave contributions that could be factorised out. The factorisation was
made possible since the only function that depends on the respective
directions of momenta was, {\it a priori}, the square of the
transition amplitudes. Indeed, it can be shown, either in the Durca or
in the Murca case, that it is a proper approximation to use some
suitably defined angle averaged values (Friman \& Maxwell \cite{fm79},
Yakovlev \& Levenfish \cite{yl95}) of $|M_{if}|^2$, instead of doing
the full angular integral. However, this approximation can be no
longer sufficient to factorise the angular integrals if some nucleons
are superfluid. This results from the fact that the Cooper pairing can
occur in anisotropic states, a feature that is explained in the next
section dealing with nucleon superfluidity in NS cores.

%%%%%%%%%%%%%%%%%%%%%%%%%%%%%%%%%%%%%%%%%%%%%%%%%%%%%%%%%%%%%%
\section{Superfluidity of nucleons}
\label{sect:sup.nuc}
%%%%%%%%%%%%%%%%%%%%%%%%%%

 Superfluidity of nucleons in NS cores is reviewed by Lombardo \&
Schulze (\cite{ls2001}) and we shall only give here a brief summary of
the results useful in the following. Neutrons are believed to form
Cooper pairs due to their interaction in the triplet $^3{\rm P}_2$
state, while protons form singlet $^1{\rm S}_0$ pairs. In the study of
the triplet-state neutron pairing, one should distinguish the cases of
the different possible projections $m_J$ of $nn$-pair angular momentum
${\vec J}$ onto a quantization axis $z$ (see, e.g., Amundsen and
{\O}stgaard \cite{ao85}): $|m_J|=0,\, 1,\, 2$. The actual
(energetically most favorable) state of $nn$-pairs is not known, being
extremely sensitive to the (still unknown) details of $nn$
interaction. One cannot exclude that this state varies with density
and is a superposition of states with different $m_J$.\\

 Hence, in the following, we shall conform ourselves to what is
usually done in the community (see, e.g., Yakovlev \etl \cite{ykgh01})
and deal with three different superfluidity types: $^1$S$_0$,
\mbox{$^3$P$_2$ ($m_J=0$)} and \mbox{$^3$P$_2$ ($|m_J|=2$)}, denoted
as A, B and C, respectively (Table \ref{tab:ABC}). The superfluidity
of type A is attributed to protons, while types B and C may be
attributed to neutrons and are sufficient to have a general idea about
possible impact of neutrons superfluidity since in these two cases its
effect are qualitatively different.\\

In addition, it can be shown (see, e.g., Yakovlev \etl \cite{ykgh01})
that in our context, it is sufficient, from the microscopic point of
view, to consider as the only effect of superfluidity the introduction
of an energy gap $\delta$ in momentum dependence of the nucleon
energy, \mbox{$\varepsilon( {\vec p} )$}. Near the Fermi level
\mbox{($ | p-p_{\rm F} | \ll p_{\rm F}$)}, this dependence can be
written as
\begin{equation} \label{eq:Gap}
\begin{array}{l}
      \varepsilon\,=\,\mu\,-\,\sqrt{\delta^2 + v_{\rm F}^2(p-p_{\rm F})^2}
      \; \; {\rm at} \; \; p < p_{\rm F} \, ,
      \\
      \varepsilon = \mu + \sqrt{\delta^2 + v_{\rm F}^2(p-p_{\rm F})^2}
      \; \; {\rm at} \; \; p \ge p_{\rm F} \; ,
\end{array}
\end{equation}
where $p_{\rm F}$ and $v_{\rm F}$ are the Fermi momentum and Fermi
velocity of the nucleon, respectively, while $\mu $ is the nucleon
chemical potential. One has \mbox{$\delta^2\,=\,\Delta^2
(T)\,F(\vartheta)$}, where $ \Delta(T)$ is the part of the gap's
amplitude that depends on the temperature, and $F(\vartheta)$
specifies dependence of the gap on the angle $\vartheta$ between the
particle momentum and the $z$ axis (Table \ref{tab:ABC}). In case A
the gap is isotropic, and $\delta = \Delta(T)$. In cases B and C, the
gap depends on $\vartheta$. Note that in case C the gap vanishes at
the poles of the Fermi sphere at any temperature: \mbox{$ F_{\rm C}
(0) = F_{\rm C} (\pi) = 0$}.\\

Notice also that the gap amplitude, $\Delta(T)$, is derived from the
standard equation of the BCS theory (see, e.g., Yakovlev et al.\
\cite{yls99}), with the value of $\Delta(0)$ that determines the
critical temperature $T_c$. The values of \mbox{$k_{\rm B} T_c /
\Delta(0)$} for cases A, B and C are given in Table \ref{tab:ABC}.\\

\begin{table}[t]
\renewcommand{\arraystretch}{1.2}
\caption{Three types of superfluidity. From Haensel \etl (\cite{hly00}) with the kind permission of the authors.}
\begin{center}
  \begin{tabular}{||c|ccc||}
  \hline \hline
Type\rule{0em}{2.5ex}
         & Pairing state &   $F(\vartheta)$
         & $k_{\rm B} T_c/\Delta(0)$  \\
  \hline
  A       & $^1{\rm S}_0$    &  1                     & 0.5669
                                                        \rule{0em}{3ex} \\
  B       & $^3{\rm P}_2\ (m_J =0)$
                             & $(1+3\cos^2\vartheta)$ & 0.8416            \\
  C       & $^3{\rm P}_2\ (|m_J| =2)$
                             & $\sin^2 \vartheta$     & 0.4926
                                        %$\displaystyle\vph{a\over F_F}$
                                        \\
  \hline \hline
\end{tabular}
\label{tab:ABC}
\end{center}
\end{table}

For further analysis it is convenient to introduce the
dimensionless quantities:
\begin{equation}
     v = \frac{\Delta(T)}{k_{\rm B}
 T}, \quad
     \tau = \frac{T}{T_c}, \quad
     y = \frac{\delta}{k_{\rm B}
 T}.
\label{DimLess2}
\end{equation}
The dimensionless gap $y$ can be presented in the form:
\begin{equation}
     y_{\rm A} = v_{\rm A}, \quad y_{\rm B} =
     v_{\rm B} \, \sqrt{1+3\cos ^2\vartheta},
     \quad
     y_{\rm C} = v_{\rm C} \, \sin \vartheta\,
\label{y}
\end{equation}
where the dimensionless gap amplitude $v$ depends only on $\tau$. In
case A the quantity $v$ coincides with the isotropic dimensionless
gap, while in cases B and C it represents, respectively, the minimum
and maximum gap (as a function of $\vartheta$) on the nucleon Fermi
surface. As we shall see later, this implies that for global
integrated quantities (such as rates), with a given value of $v$, one
can expect a stronger effect of superfluidity for the case B and a
weaker for the case C, while the case A should be in between. Notice
finally that the dependence of $v$ on $\tau$ was fitted by Levenfish
\& Yakovlev (\cite{ly94}). However, their fits will not be used in the
following. Indeed, we shall only deal with microscopic calculations
and not with astrophysical applications, and we shall therefore use
$v$ as a free parameter.

%%%%%%%%%%%%%%%%%%%%%%%%%%%%%%%%%%%%%%%%%%%%%%%%%%%%%%%%%%%%%%
\section{Out of equilibrium beta processes with superfluidity}
\label{sect:R.np}
%%%%%%%%%%%%%%%%%%%%%%%%%%%%%%%%%%%%%%%%%%%%%%
\subsection{General features of the effect of superfluidity on reactions rates} \label{sect:outgen}

 The influence of superfluidity on the beta relaxation time was
already investigated by Reisenegger (\cite{reisenegger97}), but only
in a very rough way. The principle that he used is based on what
was explained in Section \ref{sect:sup.nuc}, \ie{}, on the fact that to
estimate this influence on averaged physical quantities, such as
emissivity or reaction rates, it is sufficient, at the first level of
approximation, to only consider the creation, near the Fermi surface,
of a gap in the dispersion relation $\eps_i=\eps_i(p)$. This property can be demonstrated using Bogoliubov transformations
(see, e.g., Yakovlev \etl \cite{ykgh01}) and has, as a first obvious
consequence, strong decrease of the available
phase-space. Hence, as Reisenegger concluded, this can result in a
blocking of the reactions during quite long times, even if chemical
equilibrium is broken. Nevertheless, there are two shortcomings in the
assumption of a completely frozen composition, as made by
Reisenegger. First, the temperature is not exactly null, which implies
the existence of excited states and, therefore, of some reactions. Second,
the Cooper pairing does not always occur in a singlet spin state,
which has for consequence a possible anisotropy of the gap and of the
Fermi surface, allowing for reactions in some directions of the momentum
space.\\

 Thus, to improve the study of Reisenegger (\cite{reisenegger97}), we
 shall apply in the following the procedure that consists in adding a
 gap close to the Fermi surface in the dispersion relation, but do the
 calculations of relaxation rates through integrals similar to those
 described in Sect. \ref{sect:nonsup.beta}. In practice, it means, for
 superfluid nucleons, to replace in the integrals the usual
 dimensionless variables $x_i$ by $z_i$, with
\begin{equation} \label{Eq:zdef}
    z_i\,\widehat{=}\,\frac{\eps_i-\mu_i}{T}\,=\,{\rm sign}(x_i)\,\sqrt{x_i^2+y_i^2}\,,
\end{equation}
where 
\begin{equation} \label{Eq:xdef2}
x_i\,\widehat{=}\,\frac{v_{\F i}\,(p_i\,-\,p_{\F i})}{T}\quad{\rm and}\quad
y_i\,\widehat{=}\,\frac{\delta_i(T,\vartheta)}{T}\,\widehat{=}\,
\,v_i(T)\,{\rm F_i}(\vartheta)\,.
\end{equation}

In these expressions,
\begin{itemize}
\item [-] $v_{\F i}\,=\,(\partial \eps_i / \partial
p_i)_{p_i=p_{\F i}}$ is the Fermi velocity, making the definitions for
$x_i$, Eq.(\ref{Eq:xdef1}) and Eq.(\ref{Eq:xdef2}), equivalent near the
Fermi surface when there is no gap;
\item [-] $\delta_i(T,\vartheta)$ is the superfluid gap with $T$ the
temperature and $\vartheta$ the angle between the particle momentum and the
quantization axis (see Sect.\ \ref{sect:sup.nuc});
\item [-] $v_i(T)$ is the dimensionless gap amplitude function and
${\rm F_i}(\vartheta)$ the dimensionless anisotropy function with, for
the isotropic case (A), ${\rm F_i}(\vartheta)\,\equiv\,1$ (see
Sect.\ \ref{sect:sup.nuc}).
\end{itemize}

 The next issue that needs to be clarified before doing any calculation is
 the identification of the relevant cases to deal with. For instance,
 as it is well-known that protons can only pair in spin-singlets,
 there is no need to evaluate the effect of two simultaneous
 anisotropic pairings. Furthermore, a first approximation, which
 should not be too bad, is to assess that there is always only one type
 of superfluid nucleons, which is equivalent to assume that the
 dominant gap prevails. Hence, we shall in the following only study cases
 with superfluidity of one type of nucleons, beginning with Durca
 processes that involves less nucleons than Murca processes. Finally,
 notice that even if our goal is to calculate the superfluid
 equivalent of the rates (\ref{Eq:detgamdint}) and
 (\ref{Eq:detgammint}), we are able, with the phase-space
 approximation, to extract from these rates all microphysical
 uncertainties ($\Delta \Gamma_{D_0}$ and $\Delta \Gamma_{M_0}$),
 leaving dimensionless integrals [$I_D(\xi)$ and $I_M(\xi)$]. As a
 consequence, in order to also minimize, in the superfluid cases, the
 uncertainties coming from microphysics, it is more relevant not to
 directly calculate the superfluid rates, but the ratios between their
 values with and without superfluidity. In such a way, only
 uncertainties linked with the gap still affect the phase-space
 integrals and the results.\\

  This leads us to write, for all type of reactions, equations like
\begin{equation} \label{Eq:detgamsup}
\Delta \Gamma_X^i\,  =\,  \Delta \Gamma_{X_0}\, I_X^i(\xi,v_j)\,
=\, \Delta \Gamma_X\,{\rm R_X^i}(\xi,v_j)\,,
\end{equation}
where \mbox{${\rm
R_X^i}(\xi,v_j)\,\widehat{=}\,I_X^i(\xi,v_j)/I_X(\xi)$} are the
quantities we shall estimate: the reduction factors for the $X$ type
of reactions. In this expression, $i$ labels the type of superfluidity
(A, B or C), and $j$ the type of nucleon that is superfluid ($n$ or
$p$) with a gap $v_j$ [see definition in
Eq.(\ref{Eq:xdef2})]. Finally, notice that by definition, for $T$
greater than both critical temperatures, one has \mbox{${\rm
R_X^i}(\xi,v_j)\, =\, {\rm R_X^i}(\xi,0)\, =\, 1$}, and for $T$
smaller than at least one critical temperature \mbox{${\rm
R_X^i}(\xi,v_j)\,<\,1$}.

\subsection{Durca}

 Taking into account the fact that in the Durca case with only one
 type of superfluidity there might be only one type of
 anisotropy induced by pairing, we can write the reduction factor
 ${\rm R_D^i}(\xi,v_j)$ as
\begin{align} \label{Eq:ang.dur}
{\rm R_D^i}(\xi,v_j) & = \, \frac{1}{I_D(\xi)} \int \frac{\d \Omega}{4\pi}
\,{\rm H_D^i}(\xi,v_j)\\
 & = \frac{1}{I_D(\xi)} \int\limits_0\limits^{\pi/2} \!\! \d \vartheta \, \sin(\vartheta)
\,{\rm H_D^i}(\xi,v_j)\,,\nonumber
\end{align}
where, for type $A$ of superfluidity, ${\rm H_D^i}(\xi,v_j)$ does not depend on the
$\vartheta$ angle between the momentum of the Cooper pair and the
quantization axis.\\

  Following Eq.(\ref{Eq:ID}), one then has
\begin{multline} \label{Eq:IDS}
{\rm H_D^i}(\xi,v_j)\, \widehat{=} \, \int\limits_0\limits^\infty \d x_\nu \; x_\nu^2
       \int\limits_{-\infty}\limits^{+\infty} \d x_n\; \d x_p\; \d x_e \\
\times \{f_n\,(1-f_p)\,(1-f_e)\,\delta \left(z_n-z_p-x_e-x_\nu+\xi \right)
\\
 - f_p\,f_e\,(1-f_n)\,\delta \left(z_n+x_\nu-z_p-x_e+\xi \right)\}\,,
\end{multline}
where the $z$ variable for the non-superfluid nucleon has to be
replaced with the corresponding $x$ variable, while $v_j$ is
``included'' in the $z$ variable for the superfluid nucleon and
\mbox{$f_p\,=\,{\rm f}(z_p)\,=\,(1+e^{z_p})^{-1}$} with a similar
definition for $f_n$.\\

The integration over the electron variable can easily be done due to
the Dirac functions, and then, using the well-known formula for Fermi
integrals,
\begin{equation} \label{Eq:wk}
\int\limits_{-\infty}\limits^{+\infty} \!\!\! \d x\,\,{\rm f}(x)\,\,
{\rm f}(y-x)\,=\,\frac{y}{e^y-1}\,\widehat{=}\,{\rm G}(y)\,,
\end{equation}
enables us to integrate over the $x$ variable for the non-superfluid nucleon, giving
\begin{multline} \label{Eq:cut}
{\rm H_D^i}(\xi,v_j)\, = \, \int\limits_0\limits^\infty \d x_\nu \; x_\nu^2
       \int\limits_0\limits^{+\infty} \d x_j \\
\times \{{\rm f}(z_j)\,{\rm G}(x_\nu-\xi-z_j)+{\rm f}(-z_j)\,{\rm G}
(x_\nu-\xi+z_j)\\
-{\rm f}(z_j)\,{\rm G}(x_\nu+\xi-z_j)-{\rm f}(-z_j)\,{\rm G}(x_\nu+\xi+z_j)\}\,.
\end{multline}

This formula is the simplest that can be reached analytically and will
be evaluated with numerical methods (see Section \ref{sect:R.results}).

\subsection{Murca}

Whatever the situation with superfluidity, the main change between Durca
and Murca reactions is the presence of spectator nucleons that make the
calculations more difficult. Nevertheless, a great difference is that, in the
superfluid case, the additional dimensions of the integral no longer lead to
calculations that can quite easily be done analytically, mainly due to anisotropies.\\

Anyway, with the same method as previously described,
Eq.(\ref{Eq:detgammint}) is now replaced with
\begin{equation} \label{Eq:detgamsupmpint}
\Delta \Gamma_{M_X}^i\,=\,  \Delta \Gamma_{M_{X0}}\, I_{M_X}^i(\xi,v_j)
= \Delta \Gamma_{M_X}\,{\rm R_{M_X}^i}(\xi,v_j)\,,
\end{equation}
with ${\rm R_{M_X}^i}(\xi,v_j)\,=\,I_{M_X}^i(\xi,v_j)/I_M(\xi)$, while
$X$ is the index of the branch ($n$ or $p$), $i$ the index of the
superfluidity type and $j$ the index of the superfluid nucleon. Here
again, for $T$ greater than the critical temperature, \mbox{one has
${\rm R_{M_X}^i}(\xi,v_j)\,=\,{\rm R_{M_X}^i}(\xi,0)\,=\,1$} and in
other cases \mbox{${\rm R_{M_X}^i}(\xi,v_j)\,<\,1$}.\\

Notice that $I_{M_X}^i(\xi,v_j)$ has to include some angular factor
$A_{M_X}$, similar to $\d \Omega/(4 \pi)$ in Eq.(\ref{Eq:ang.dur}), but
somehow more complicated. Hence, we have
\begin{multline} \label{Eq:IMS}
   I_{M_X}^i(\xi,v_j)\,=\, \frac{4\pi}{A_{M_X}} \int \prod_{j=1}^5 \d \Omega_j 
          \int\limits_0\limits^{\infty} \d x_\nu \, x_\nu^2 
          \left[ \prod_{j=1}^5 \int\limits_{-\infty}\limits^{+\infty}
          \d x_j \, f_j \right] \\
 \times\,\delta \left( \sum_{j=1}^5 {\vec{p}}_j \right)\,
\left[ \delta \left( x_\nu - \xi - \sum_{j=1}^5 z_j \right) - 
\delta \left( x_\nu +\xi - \sum_{j=1}^5 z_j \right) \right]
\end{multline}
where we have introduced a condensed notation for all particles, with
exception of the neutrinos : $j$ is the index of the particles with
$j=1\,..\,5$, respectively, corresponding to $n, p, N_i, N_f, e$. With
this notation, $\prod\limits_{j=1}\limits^5 \d \Omega_j$ is a global
angular element of integration, while all other notations ($z_j$,
$f_j$) agree with previously defined notations with the convention
that for a non-superfluid nucleon or an electron
$z_j\,\equiv\,x_j$. Finally, notice that the $4 \pi$ factor in front of
the expression comes from the integration over possible directions of
the neutrino momentum.\\

To complete this section, we will give more advanced analytical
formulae corresponding to some specific cases we will deal with in the
following, not only because they are the easiest to treat numerically,
but also because it is sufficient due to the still huge error-bars
concerning dense matter microphysics. Those cases are
\begin{itemize}
\item [-] the neutron branch with isotropic superfluidity of protons (gap A);
\item [-] the proton branch with any superfluidity of neutrons;
\item [-] the neutron branch with isotropic superfluidity of neutrons (gap A).
\end{itemize}

In the first case, due to full isotropy, one can write
\begin{equation} \label{Eq:IMS2}
   I_{M_n}^{pA}(\xi,v_p)  =  \frac{1}{g(\xi)}\,
 \int\limits_{-\infty}\limits^{+\infty} \d x_p \,{\rm f}(z_p)\,\left({\rm H}(z_p+\xi)
-{\rm H}(z_p-\xi)\right)\,,
\end{equation}
with
\begin{equation}
{\rm H}(x) \,\widehat{=}\,  \int\limits_0\limits^{+\infty}\d s \, s^2 \frac{s-x}{\exp(s-x)-1}
\,\left((s-x)^2\,+4\pi^2\right)\,
\end{equation}
and
\begin{equation}
g(\xi)\,\widehat{=}\,\xi\,\frac{367\,\pi^6}{252}\left(1\,+\,\frac{189\,\xi^2}{367\,\pi^2}
+\,\frac{21\,\xi^4}{367\,\pi^4}+\,\frac{3\,\xi^6}{1835\,\pi^6}\right)\,.
\end{equation}

 In the second case, one has
\begin{multline} \label{Eq:IMS3}
   I_{M_p}^{nX}(\xi,v_n)  = \frac{1}{g(\xi)}\,
 \int\limits_0\limits^1 \d \mathfrak{c} \int\limits_{-\infty}\limits^{+\infty} \d x_n \,{\rm f}(z_n)\\
\times\,\left({\rm H}(z_n+\xi)-{\rm H}(z_n-\xi)\right)\,,
\end{multline}
where $\mathfrak{c} \widehat{=} \cos(\vartheta)$ appears in the $y_n$ variable
while $g(\xi)$ and $H(x)$ are the same as in Eq.(\ref{Eq:IMS2}).\\

 The third case leads to an integral to evaluate numerically that is
 more complicated, since analytical integration can be performed only
 on the electron and non-superfluid nucleon variables. With the usual
 technique, one gets
\begin{multline} \label{Eq:IMS4}
   I_{M_n}^{nA}(\xi,v_n) = \frac{1}{I_M(\xi)} \int\limits_{-\infty}\limits^{+\infty} \d x_{n1}\,\d x_{n2}\,\d
 x_{n3} \,{\rm f}(z_{n1})\,{\rm f}(z_{n2})\,{\rm f}(z_{n3})\,\\
 \left({\cal H}(z_{n1}+z_{n2}+z_{n3}+\xi)-{\cal H}(z_{n1}+z_{n2}+z_{n3}-\xi)\right)\,,
\end{multline}
with
\begin{equation}
{\cal H}(x) \,\widehat{=}\,  \int\limits_{-\infty}\limits^{+\infty}\d s \, s^2 \frac{s-x}{\exp(s-x)-1}\,,
\end{equation}
and where the normalisation function $I_M(\xi)$ was defined in
Eq.(\ref{Eq:IM3}).\\

 For the numerical calculation, as was done for Durca reactions [see
Eq.(\ref{Eq:cut})], all those integrals for Murca reactions are cut in
pieces in such a way that the $x_i$ variables take only positive
values. However, we shall not give here more details on the algorithms
that we used, and we send the reader to the Appendix or to the
subroutine comments on
\mbox{``http://luth2.obspm.fr/\~{}etu/villain/Micro/Reduction.html''}. Instead, the
next section will now describe the main numerical results.

%%%%%%%%%%%%%%%%%%%%%%%%%%%%%%%%%%%%%%%%%%%%%%%
\section{Numerical results}
\label{sect:R.results} 
%%%%%%%%%%%%%%%%%%%%%%%%%%

 Due to the complexity of the multidimensional integrations, we
 decided to test several numerical integration algorithms (see the
 Appendix for more details) and finally we obtained results in perfect
 agreement in the physical parameter range of interest. Here, we shall just give
 a graphical overview of those results for
\begin{description}
\item[-] Durca reaction with superfluidity of each type;
\item[-] Murca reaction with isotropic superfluidity of the
non-spectator nucleon (e.g. proton for the neutron branch);
\item[-] Murca reaction with isotropic superfluidity of the spectator
nucleon (e.g. neutron for the neutron branch).
\end{description}

\begin{figure}
   \centering
   \includegraphics[width=0.5\textwidth]{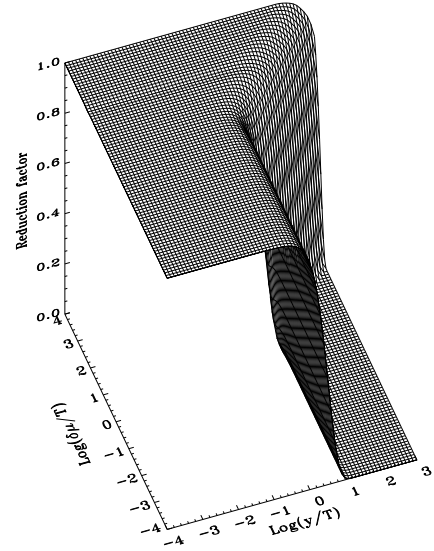}
      \caption{Three dimensional view of the reduction factor for
      Durca reaction with isotropic superfluidity of the protons. Remember that here by $T$ we mean $k_{\rm b}\,T$ due to the chosen units.}
         \label{3D}
   \end{figure}

  The first graph (Fig. \ref{3D}) depicts in a three-dimensional way
  the reduction factor for the simplest situation, which is Durca
  reaction with isotropic superfluidity (of protons). We shall not
  give more 3D figures like this one, because they are not the best to
  look at the influence of the type of superfluidity, even if they
  help to have a quick idea of the effect of superfluidity. Indeed,
  this figure shows that for huge departures from equilibrium,
  superfluidity really has an impact only when the amplitude of the
  gap is of the order of $\delta \mu$, or is much larger than $\delta
  \mu$. In the first case, there is an exponential decay of the
  reduction factor, which leads, for the second case, to the expected
  ``frozen composition'' (reduction factor equal to $0$). Yet, for
  very small values of the gap's amplitude (in respect to the
  temperature), the ratio with $\delta \mu$ does not matter and
  superfluidity is not relevant since the reduction factor is very close to $1$.\\

  \begin{figure}
   \centering
   \includegraphics[angle=-90,width=0.5\textwidth]{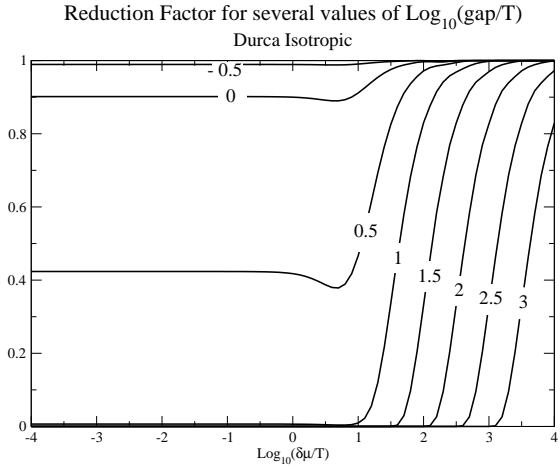}
      \caption{Reduction factor as a function of the departure from
      beta equilibrium for several values of the gap's
      amplitude. Durca reaction with isotropic superfluidity of the
      protons. Remember that here by $T$ we mean $k_{\rm b}\,T$ due to the chosen units.}
         \label{DurcaISo}
   \end{figure}

\begin{figure}
   \centering
   \includegraphics[angle=-90,width=0.5\textwidth]{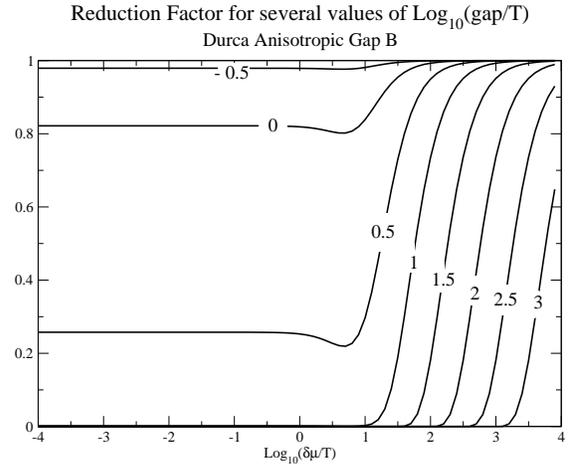}
 \caption{Reduction factor as a function of the departure from beta
 equilibrium for several values of the gap's amplitude. Durca reaction
 with anisotropic superfluidity type B. Notation as in \mbox{Fig. \ref{DurcaISo}}.}
         \label{DurcaB}
   \end{figure}

\begin{figure}
   \centering
   \includegraphics[angle=-90,width=0.5\textwidth]{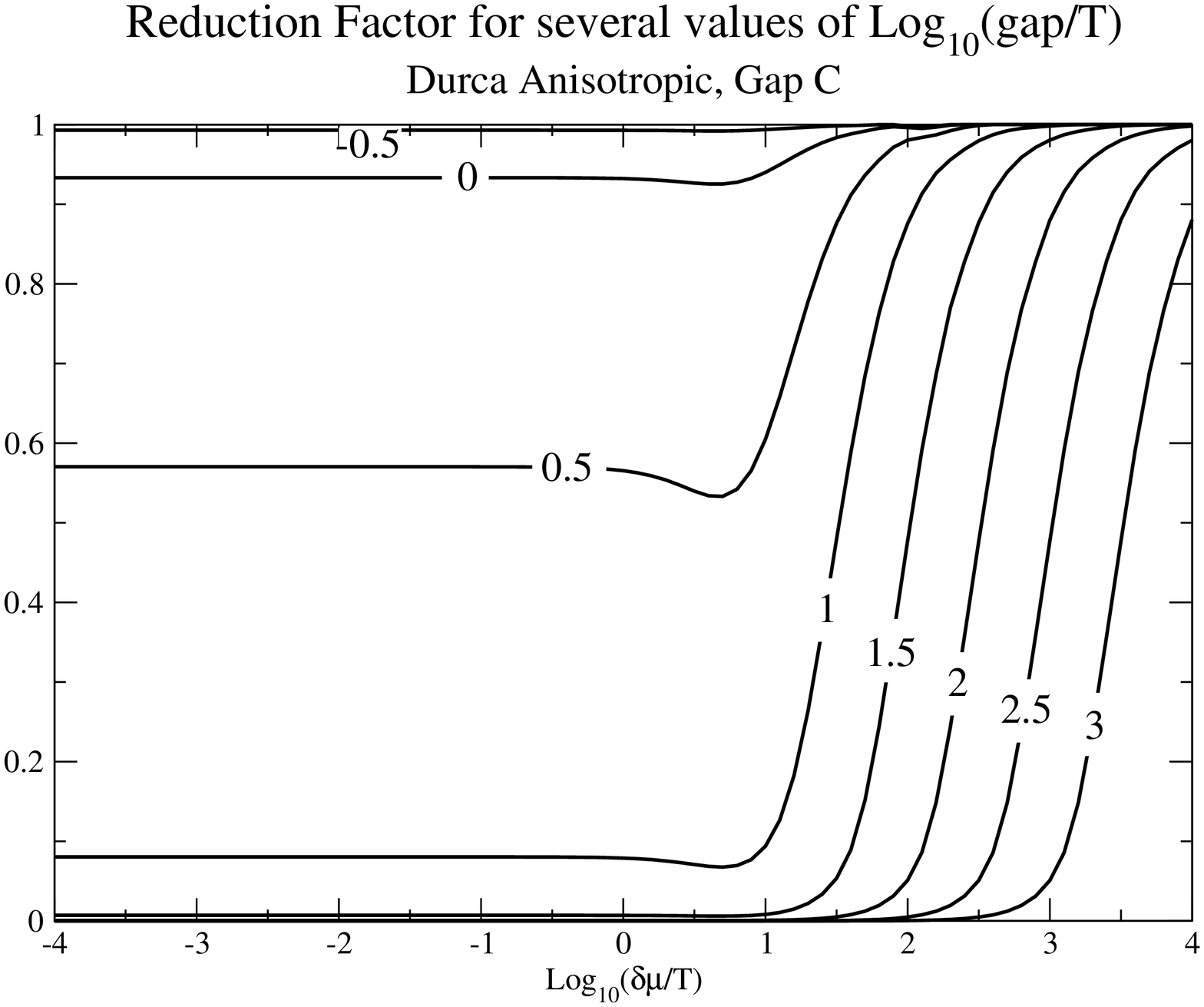}
  \caption{Reduction factor as a function of the departure from beta
  equilibrium for several values of the gap's amplitude. Durca
  reaction with anisotropic superfluidity type C. Notation as in \mbox{Fig. \ref{DurcaISo}}.}
         \label{DurcaC}
   \end{figure}

\begin{figure}
   \centering
   \includegraphics[angle=-90,width=0.5\textwidth]{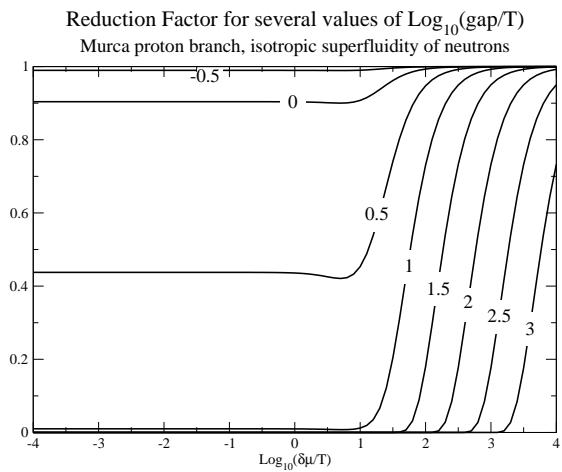}
 \caption{Reduction factor as a function of the departure from beta
 equilibrium for several values of the gap's amplitude. Murca reaction
 with isotropic superfluidity of one single nucleon. Notation as in \mbox{Fig. \ref{DurcaISo}}.}
         \label{Murca1}
   \end{figure}

\begin{figure}
   \centering
   \includegraphics[angle=-90,width=0.5\textwidth]{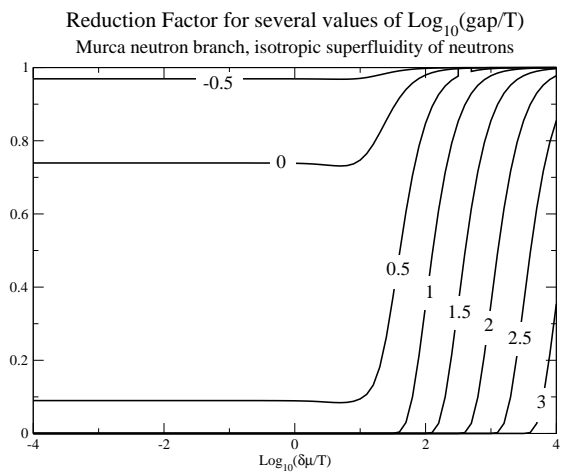}
 \caption{Reduction factor as a function of the departure from beta
 equilibrium for several values of the gap's amplitude. Murca reaction
 with isotropic superfluidity of three nucleons out of four which are
 involved. Notation as in \mbox{Fig. \ref{DurcaISo}}.}
         \label{Murca2}
   \end{figure}

  Fig. \ref{DurcaISo} shows exactly the same results, but in another
  way. Here, the variable is $\delta \mu$, and each curve corresponds
  to a given value of the ratio $gap$/$temperature$. For huge gaps,
  $Log_{10}(gap/T)\,\ge\,1$, it can be noticed that as soon as $\delta
  \mu\,\ge\,gap$, and until the reduction factor becomes close to $1$,
  the curves are parallel. The meaning of this feature of the graphs
  is that, at least for that part of the curves, some functions $\mathfrak{f}$ and
  $\mathfrak{g}$ exist, such that the reduction factor can be written as
  $R\,\sim\,\mathfrak{f}(gap)\,\mathfrak{g}(gap/\delta \mu)\,.$
  Furthermore, comparison of Figs. \ref{DurcaB} and \ref{DurcaC} with
  Fig. \ref{DurcaISo} enables to see that the effect of anisotropy of
  the gap cannot easily be evaluated since it can either increase
  (type C) or decrease (type B) the reduction factor in respect to its
  value in the isotropic case. However, what seems to be the most
  important is the maximal value of the gap on the Fermi surface since
  the value for case B is smaller than for case A, while for case C it
  is larger. As far as Murca reactions are concerned
  (Figs. \ref{Murca1} and \ref{Murca2}), the situation is the
  following. If only one nucleon (the non-spectator one) is
  superfluid, then the difference with the Durca case seems very small
  and almost not visible (see Figs. \ref{DurcaISo} and
  \ref{Murca1}). However, if the spectator nucleons (the most
  represented) are also superfluid the impact is much larger, as is
  depicted by Fig. \ref{Murca2}. Nevertheless, it should not be
  forgotten that the reduction factors are calculated for relaxation
  times that are really different : Durca reactions allow for a much
  faster relaxation towards equilibrium (see, e.g., sect. 3.5 of
  Yakovlev \etl \cite{ykgh01} and references therein).

%%%%%%%%%%%%%%%%%%%%%%%%%%%%%%%%%%%%%%%%%%%%%%%
\section{Discussion and conclusions}
\label{sect:disc.concl}
%%%%%%%%%%%%%%%%%%%%%%%%%%

We calculated the effect of nucleon superfluidity on the rates of beta
processes in neutron star matter which is off beta
equilibrium. Superfluidity reduces these rates by a factor that
depends on the types of the beta process (direct Urca or modified
Urca) and of superfluidity, but also on the dimensionless gaps and
chemical-potential mismatch parameters $\Delta/k_{\rm B}T$ and
$\delta\mu/k_{\rm B}T$, respectively. Due to the degeneracy of nuclear
matter, the reduction factors do not depend explicitly on the EOS (see
Sect. \ref{sect:outgen}).\\

In order to reduce the number of parameters and the technical
difficulties in calculating the integrals to evaluate, we assumed that
we could always consider only one type of nucleon being superfluid,
the larger gap prevailing. Such an approximation, which seems to be
quite reasonable for the nuclear matter of NS cores (see Lombardo \&
Schulze \cite{ls2001}), indeed allows to analytically integrate more
variables (see Sect. \ref{sect:R.np}), strongly reducing the computational time needed.\\

In the case of the direct Urca process, reduction factors values were
calculated for the isotropic spin-singlet $^1{\rm S}_0$ pairing of
protons and for two types of the angle-dependent gaps resulting from
the spin-triplet $^3{\rm P}_2$ pairing of neutrons. Since, in the
modified Urca case, the phase-space integrals, which can be of up to
the $12^{\rm th}$ order, are very complex, we decided to first neglect
the superfluid gaps anisotropy. This approximation can seems quite
rough, however, it is supported by the fact that anisotropy of the gap
can either increase or decrease the effect of superfluidity in respect
to the case of the isotropic gap (see Sect. \ref{sect:R.results}),
while the actual state of the triplet neutron Cooper pairs is still poorly known
(see Sect. \ref{sect:sup.nuc}).\\

Subroutines for the numerical calculation of the superfluid reduction
factors are briefly described in the Appendix with link to the website
on which they can be found. They yield the reduction factors for given
$\Delta/k_{\rm B}T$ and $\delta\mu/k_{\rm B}T$. One has then to
multiply by these factors the analytical expressions for the
non-superfluid beta reaction rates to get the physical rates for
superfluid matter.\\

Our results, combined with an equation of state of the neutron star
core, can be used in numerical simulations of neutron star pulsations,
to account for the dissipation due to the non-equilibrium beta
processes implied by the local baryon density variations. Such a
modeling will be valid both in a highly nonlinear suprathermal regime
of dissipation $\delta\mu>k_{\rm B}T$ as well as in the $\delta\mu \ll
k_{\rm B}T$ limit where the bulk viscosity description can be used
(see Haensel \etl \cite{hly00}, \cite{hly01}, \cite{hly02}). Another
application concerns the calculation of non-equilibrium heating and
neutrino emission of superfluid matter in a spinning down pulsar,
which would allow one to model in a realistic manner the
cooling/slowing-down of superfluid pulsars. Both problems will be
studied by us in the near future.\\

The formulae presented in the present article were obtained under the
simplifying assumption that neutrons, protons, and electrons move
together as one single fluid, \ie, they are all at rest in one single
reference system (local rest-frame). On a dynamic timescale, such an
approximation is correct if all three fluids are normal (see, e.g.,
Villain \etl \cite{villain05}). However, in the most general case of
superfluid neutron and protons, we have to deal with three fluids, and
therefore with three different flows of matter. First, we have a
normal fluid composed of electrons and of normal components of
neutrons and protons (this is a fluid of excited neutron and proton
quasi-particles). Then, we have two flows of superfluid condensates,
the neutron and proton one, connected between themselves by the
superfluid entrainment (Andreev-Bashkin effect). If we connect the
local rest-frame with normal fluid, then the equilibrium distributions
of neutron and proton quasiparticles in the condensates should be
calculated taking into account the relative velocities
$\vec{u}_{ne}={\vec u}_{n}-{\vec u}_{e}$ and $\vec{u}_{pe}={\vec
u}_{p}-{\vec u}_{e}$, respectively. For example, for isotropic gaps,
in the rest-frame comoving with normal fluid, the neutron superfluid
quasiparticle energies are
$\varepsilon_n^\prime(\vec{p})=\varepsilon_n(\vec{p}-m\vec{u}_{ne})+\gamma_{nn}\,\vec{p}\cdot\vec{u}_{ne}+\gamma_{np}\,\vec{p}\cdot\vec{u}_{pe}$,
where $m$ is neutron mass, and terms quadratic in $\vec{u}_{ne}$ and
$\vec{u}_{pe}$ were neglected; the last two terms describe the
entrainment effects and $\gamma_{\alpha\beta}$ ($\alpha,
\beta\,=\,n,p$) is the microscopic entrainment matrix calculated by
Gusakov \& Haensel (\cite{gh05}). Generally, macroscopic flows
velocities are small compared to the Fermi velocities and lowest-order
approximations are valid. The calculations in the three-fluid model
will be the next step of our study of non-equilibrium beta processes
in superfluid neutron star cores.

\acknowledgement{We are grateful to K.P. Levenfish for precious
comments, and to J.-M. Chesneaux, F. J\'ez\'equel and F. Rico for
stimulating discussions. This work was partially supported by the
Polish KBN grant no. 1-P03D-008-27 and by the Associated European
Laboratory LEA Astro-PF of PAN/CNRS. LV also generously benefited from
the hospitality of the Southampton University General Relativity Group
during part of the development of this work.}

\section{Appendix: Numerical technics}

  The integrals presented in this article, which were to be evaluated
  numerically, are of quite standard type in the field of degenerate
  Fermi fluids. However, as they involve various mathematical features
  that can easily lead to numerical difficulties, we have estimated
  that it was worth interacting with some applied mathematicians
  expert in sophisticated algorithms to dynamically control the
  accuracy of numerical multi-dimensional integrals. These people,
  J.-M. Chesneaux, F. J\'ez\'equel, F. Rico and M. Charrikhi from LIP6
  of Paris 6 University, are part of the ``Cadna team'', Cadna being
  the acronym for ``Control of Accuracy and Debugging for Numerical
  Applications'' and also the name of the numerical tool they created
  (\mbox{``http://www-anp.lip6.fr/cadna/''}). We shall not give here a
  description of their algorithm, but let us just mention that it is
  based on discrete stochastic arithmetic, and allows for calculations
  with a chosen precision (see, e.g., Chesneaux \& J\'ez\'equel
  \cite{chesneaux98} and
  J\'ez\'equel \cite{jezequel04}).\\

  Yet, the drawback of this method is that the time needed to carry
  out the computation can be very long, which prohibits to directly
  use it in physical simulations. As a consequence, we will no longer
  mention the Cadna code in the following, but it shall be noticed
  that we used it to test and calibrate our less sophisticated
  algorithms, and then to get clear ideas on the number of
  numerical points needed for various values of the
  parameters (see J\'ez\'equel \etl \cite{jrcc05} and Charrikhi \etl
  \cite{charrikhi02}). Indeed, as we shall see now, the features of the
  integrals make that the needed number of points is not uniform in
  the $(v=\Delta/k_{\rm B}T\,,\xi=\delta\mu/k_{\rm B}T)$ plane.\\

 The main difficulties in the integrations to perform are
\begin{description}
\item[-] infinite size of the domains of integration;
\item[-] singularities in the denominators of some factors (e.g. \mbox{$e^x-1$} for $x=0$);
\item[-] external free parameters (dimensionless gap amplitude $v$ and
dimensionless departure from equilibrium $\xi$) that can be very large
or different.
\end{description}

 For not too precise calculations, the first of these issues is in
 fact quite easily dealt with, as using the asymptotic behaviours of
 exponential functions, one can show that it is not worth doing the
 integration on the full domain. Hence, the numerical domains are cut
 to some maximal values, which were chosen in such a way that
 exponential functions make further integration irrelevant. More
 precisely, due to the Fermi-Dirac distributions and to the
 ``G-function'' defined in Eq.(\ref{Eq:wk}), one can see that the
 typical relevant scale for the ``$x$-variable'' of a
 superfluid nucleon is \mbox{$x\,\sim\,s_x\,\equiv\,\xi\,+\,\sqrt{2\,v}$} and that
 the scale for the neutrino variable is
 \mbox{$x_{\nu}\,\sim\,s_\nu\,\equiv\,\xi\,+\,v$}. The maximal values for the
 numerical integration were 10 times the sum of 10 and of those typical scales, to
 prevent the occurence of problems for vanishing $v$ and $\xi$.\\

As far as singularities were concerned, we just dealt with them in
 putting in the code not exactly the functions but, beyond threshold
 values, some asymptotic formulae that are very good due to the
 exponential behaviours. The only trouble in this is that the presence
 of the possibly very large values of the $v$ and $\xi$ parameters can
 make the cut-off values very huge. However, the total number of
 numerical points needed was kept quite reasonable using 
\begin{description}
\item[-] some ``logarithmic variables'' (a suggestion done by
 V. Bezchastnov to K. Levenfish who kindly shared it with us):
 instead of directly integrating over the $x$-variables, we made the
 integrations over some $t$-variables defined as
\begin{equation}
x\,\equiv\,s_x\,\left(e^t\,-\,1\right)\,,
\end{equation}
where $s_x$ is the typical scale defined above;
\item[-] spectral decomposition: we evaluated the integrals with the
 so-called Gauss-Legendre quadrature method (see, e.g., Krylov
 \cite{krylov62}).
\end{description}

 In this way, the main reason for needing huge numbers of numerical
 points were either an anisotropic gap with a very large amplitude, or
 small gap amplitude but combined with large departure from beta
 equilibrium. Anyhow, the subroutines used to produce the results
 displayed in Sect. \ref{sect:R.results} are gathered on the website
 \mbox{``http://luth2.obspm.fr/\~{}etu/villain/Micro/Reduction.html''},
 together with some comments and tests, and they include the
 determination of reasonable numbers of numerical points for a wide
 range of the $v$ and $\xi$ parameters. Typically the proposed values
 (that can easily be changed playing in the subroutines) allow for a
 relative precision always at least better than $10^{-3}$ for $v$ in
 $\left[0,10^3\right]$ and $\xi$ in $\left[0,10^4\right]$. Of course,
 the subroutines can reach a much better precision just with a few
 additional numerical points and without huge change in the CPU time, but since there
 was no need for it, we limited the precision at that level.


\begin{thebibliography}{}

\bibitem[1985]{ao85}
   Amundsen~L., {\O}stgaard~E., 1985,
   Nucl.\ Phys., A442, 163


\bibitem[2001]{andersson01}
Andersson~A., Comer~G.L., 2001, Mon. Not. of the Royal Astron. Soc., 328, 1129

\bibitem[2002]{charrikhi02} Charrikhi~M., Chesneaux~J.-M.,
J\'ez\'equel~F., Rico~F., Villain~L., 2002, proceeding of the SCAN2002
conference, Paris (France), 23-27 September 2002

\bibitem[1998]{chesneaux98}
Chesneaux~J.-M., J\'ez\'equel~F., 1998, J. of Univ. Comp. Science, 4(1), 2 

\bibitem[2005]{fernandez05} Fern\'andez~R., Reisenegger, A., 2005, ApJ, 625, 291

  \bibitem[1979]{fm79} Friman B.L., Maxwell O.V., 1979, ApJ, 232, 541

\bibitem[1993]{gh93} Gourgoulhon E., Haensel P., 1993, A\&A, 271, 187

\bibitem[2005]{gh05} Gusakov M.E., Haensel P., 2005, Nucl. Phys. A, 761, 333

  \bibitem[1992]{h92} Haensel P., 1992, A\&A, 262, 131

\bibitem[2000]{hly00} Haensel P., Levenfish K.P., Yakovlev D.G., 2000, A\&A, 357, 1157

\bibitem[2001]{hly01} Haensel P., Levenfish K.P., Yakovlev D.G., 2001, A\&A, 372, 130

\bibitem[2002]{hly02} Haensel P., Levenfish K.P., Yakovlev D.G., 2002, A\&A, 381, 1080

\bibitem[2004]{jezequel04} J\'ez\'equel~F., 2004, App. Num. Maths., 50(2), 147 

\bibitem[2005]{jrcc05} J\'ez\'equel~F., Rico~F., Chesneaux~J.-M., Charrikhi~M., 2005, Math. and Comp. in Simulation, to appear. 

\bibitem[1962]{krylov62} Krylov, V. I., 1962, Approximate calculation
of integrals, ACM Monograph Series, The MacMillan Company, New York,
NY

\bibitem[1991]{lpph91}
   Lattimer~J.M., Pethick~C.J., Prakash~M., Haensel~P.,
   1991, Phys.\ Rev.\ Lett., 66, 2701

%  \bibitem[1993]{ly93} Levenfish, K. P., Yakovlev, D. G., 1993, in Strongly
%Coupled Plasma Physics, eds. H. M. Van Horn and S. Ichimaru
%(Univ. of Rochester Press, Rochester), 167

\bibitem[1995]{lee95} Lee~U., 1995, A\&A, 303, 515


\bibitem[1994]{ly94}
   Levenfish~K.P., Yakovlev~D.G., 1994, Astron.\ Lett., 20, 43

\bibitem[1995]{lm95}
   Lindblom~L., Mendel~G., 1995, ApJ, 444, 804

\bibitem[1998]{lom98}
   Lindblom~L., Owen~B.J., Morsink~S., 1998,
   Phys.\ Rev.\ Lett., 80, 4843

\bibitem[2001]{ls2001}
    Lombardo U., Schulze H.-J., 2001,
    in: Physics of Neutron Star Interiors, D. Blaschke,
    N.K. Glendenning, A. Sedrakian (Eds.) Springer,
    Heidelberg, p.30

\bibitem[2002]{prix02} Prix R., Rieutord M., 2002, A\&A, 393, 949

   \bibitem[1992]{reisenegger92} Reisenegger A., Goldreich~P., ApJ, 395, 240

   \bibitem[1995]{reisenegger95} Reisenegger, A., 1995, ApJ, 442, 749

   \bibitem[1997]{reisenegger97} Reisenegger, A., 1997, ApJ, 485, 313

   \bibitem[1983]{st83}
   Shapiro S.L., \& Teukolsky S.A., 1983,
Black Holes, White Dwarfs and Neutron Stars,
   (Wiley-Interscience, New-York)

\bibitem[1996]{UrpSha1996}
   Urpin~V.A., Shalybkov~D.A., 1996, MNRAS, 281, 145

\bibitem[2005]{villain05}
   Villain~ L., Bonazzola~S., Haensel~P., 2005, Phys.\ Rev.\ D, 71, 083001

\bibitem[1995]{yl95}
   Yakovlev~D.G., Levenfish~K.P., 1995,
   A\&A, 297, 717

\bibitem[1999]{yls99}
   Yakovlev~D.G., Levenfish~K.P., Shibanov~Yu.A., 1999,
   Physics-Uspekhi, 169, 825

   \bibitem[2001]{ykgh01} Yakovlev, D. G., Kaminker, A. D., Gnedin, O. Y., \&
Haensel, P., 2001, Phys. Rep. 354, 1.


\end{thebibliography}
\end{document}